\documentclass[twocolumn,pra,superscriptaddress,floatfix]{revtex4}
\usepackage{amsmath}
\usepackage{amssymb}
\usepackage{amsmath}
\usepackage{graphicx}

\newcommand{\CNOT}{\textsc{cnot}}

\newcommand{\SWAP}{\textsc{swap}}

\newcommand{\ket}[1]{\ensuremath{|{#1}\rangle}}
\newcommand{\bra}[1]{\ensuremath{\langle{#1}|}}
\newcommand{\nuc}[2]{\ensuremath{^{#1}\textrm{#2}}}
\newcommand{\A}{\textsc{a}}
\newcommand{\B}{\textsc{b}}
\newcommand{\C}{\textsc{c}}
\newcommand{\D}{\textsc{d}}

\begin{document}

\title{Witnesses of non-classicality for simulated hybrid quantum systems}

\author{G. Bhole}
\affiliation{Clarendon Laboratory, University of Oxford, Parks Road, Oxford OX1 3PU, United Kingdom}
\author{J. A. Jones}
\affiliation{Clarendon Laboratory, University of Oxford, Parks Road, Oxford OX1 3PU, United Kingdom}
\author{C. Marletto}
\affiliation{Clarendon Laboratory, University of Oxford, Parks Road, Oxford OX1 3PU, United Kingdom}
\affiliation{Centre for Quantum Technologies, National University of Singapore, 3 Science Drive 2, Singapore 117543}
\author{V. Vedral}
\affiliation{Clarendon Laboratory, University of Oxford, Parks Road, Oxford OX1 3PU, United Kingdom}
\affiliation{Centre for Quantum Technologies, National University of Singapore, 3 Science Drive 2, Singapore 117543}
\date{\today}

\begin{abstract}
The task of testing whether quantum theory applies to all physical systems and all scales requires considering situations where a quantum probe interacts with another system that need not obey quantum theory in full. Important examples include the cases where a quantum mass probes the gravitational field, for which a unique quantum theory of gravity does not yet exist, or a quantum field, such as light, interacts with a macroscopic system, such as a biological molecule, which may or may not obey unitary quantum theory. In this context a class of experiments has recently been proposed, where the non-classicality of a physical system that need not obey quantum theory (the gravitational field) can be tested indirectly by detecting whether or not the system is capable of entangling two quantum probes. Here we illustrate some of the subtleties of the argument, to do with the role of locality of interactions and of non-classicality, and perform proof-of-principle experiments illustrating the logic of the proposals, using a Nuclear Magnetic Resonance quantum computational platform with four qubits.
\end{abstract}

\maketitle

A recently proposed class of experiments has brought the possibility of testing quantum effects in gravity closer to current experimental capabilities \cite{Marletto2017,Bose2017}. The remarkable feature of these experiments is that they are based on a general argument, whereby if an intermediate system (which need not obey quantum theory) can mediate entanglement between two quantum systems, then it itself must be non-classical. By a system being non-classical we mean, following \cite{Marletto2017}, that the system has at least two non-commuting variables. This is a weaker property than displaying full quantum coherence: it means, operationally, that the system has at least two variables with the property that they cannot be measured simultaneously to an arbitrarily high accuracy. This is a remarkably general argument, which can be applied to any mediator whatever its physical origin happens to be. It therefore generalises theoretical considerations \cite{Marletto2017a, Marletto2017b}, which date back to Feynman's and DeWitt's arguments for the quantisation of gravity, aiming at hybrid systems (those composed of a quantum probe system, that obeys quantum theory, and another system whose dynamics and scale are not fully specified). The argument sets a novel paradigm which will be crucial for the exploration of tests beyond currently known dynamical laws, specifically to witness non-classicality in systems that may not obey quantum theory.

In preparation for an actual experiment involving superposed masses interacting through gravity, in this paper we intend to further clarify the logic of the argument, using a quantum simulation. Specifically we shall illustrate how the degree of non-commutativity of relevant variables of the entanglement mediator relates to the final entanglement of the probes, in a specific quantum model, and the important role of locality of interactions in the argument.

Based on this, we also propose an experimental simulation using four Nuclear Magnetic Resonance (NMR) qubits arranged in a linear chain. The local transfer of entanglement takes place from one end of the chain to the other, through a mediator which may be non-classical, in the sense that it may have a pair of non-commuting observables. In this simulation, the mediator is the third qubit in the chain, which can either be undisturbed or undergo dephasing, which simulates the classical limit in which only one observable of the qubit can be accessed, effectively making it behave like a classical bit. We show that mediated entanglement disappears in the presence of complete dephasing, which corresponds to the mediator behaving classically, i.e.\ not being able to access at least two non-commuting degrees of freedom.

\section{Quantum gate model}

The idea of the test of non-classicality is elegant and can be illustrated via a quantum simulation, as follows.

Consider three systems: two qubits, $Q_1$ and $Q_2$, and another system $S$ (the mediator), which is only assumed to have a classical observable $T$, meaning one that can in principle be perfectly measured. This mediator system could be the gravitational field, for example, but could be more general.

Suppose that they are all initialised in a state where they are not entangled and that interactions are allowed between $Q_1$ and $S$, and between $Q_2$ and $S$, but (and this is essential) {\sl not} between $Q_1$ and $Q_2$. If at some point later in the evolution $Q_1$ and $Q_2$ become entangled, then one can infer that $S$ must have at least another variable $W$ that is complementary to $T$, meaning that $T$ and $W$ cannot be perfectly measured by the same device. Therefore $T$ and $W$ can be represented as two non-commuting degrees of freedom of $S$.

For the purpose of our simulation, the formation of entanglement through non-commuting degrees of freedom of the mediator can be modelled in a number of equivalent ways using quantum theory. In \cite{Marletto2017a} two of us proposed a Hamiltonian model in linear quantum field theory, applicable to either gravity or electromagnetism, where the system $S$ is treated as a single harmonic oscillator, while $Q_1$ and $Q_2$ are two masses that can be put into spatial superpositions of two different locations.  Here we focus instead on two quantum network models, with the aim of eventually realising a quantum simulation of the effect.

The first network is chosen so that the interaction between $Q_1$ and $Q_2$ is symmetric, mirroring the original quantum field theory Hamiltonian interaction.

To uncover the role of the non-commuting variables of the mediator $S$ in mediating entanglement, it is illuminating to resort to the Heisenberg picture for quantum information (see, for example, \cite{DEU,Gottesman1999}). The Heisenberg picture is more suitable to track the information transfer residing in non-commuting observables, which establishes entanglement. The two pictures are, of course, equivalent, but the Heisenberg one is more direct for our purposes.

Consider a chain of four qubits, A, B, C, and D.  Let $q_{x\alpha}$ denote an operator representing the $x$-component of qubit $\alpha$, and similarly for the $y$ and $z$ components. These operators act on the $2^{4}$-dimensional Hilbert space of the four qubits. We have $q_{z\alpha}q_{x\alpha}= {\rm i} q_{y\alpha}$, $q_{z\alpha}^2={id}$ and likewise for all the other components, while components of different qubits commute. If the gate $U(t_n)$ operates between time $t_n$ and $t_{n+1}$, we shall denote by
\begin{equation}
O_{\alpha}(t_{n+1})=U(t_n)^{\dagger}O_{\alpha}(t_{n})U(t_n)
\end{equation}
the operator representing the observable $O$ of system ${\alpha}$ after its action. The initial conditions are fixed by choosing particular values for $q_{x\alpha}(t_{0})$, $q_{y\alpha}(t_{0})$, $q_{z\alpha}(t_{0})$, for all $\alpha$'s, and by the Heisenberg state $\rho_H$. The state of each qubit $\alpha$ at time $t$ is completely specified by at least two components, e.g.\ $\{q_{x\alpha}(t), q_{z\alpha}(t)\}$. The state of the joint system is likewise reconstructed given all of the observables in the set $\{q_{x\alpha}(t)$, $q_{z\alpha}(t)\}$, because
\begin{equation}
U(t_{n})q_{x\alpha}(t_n)q_{z\alpha}(t_n)U^{\dagger}(t_{n})=q_{x\alpha}(t_{n+1})q_{z\alpha}(t_{n+1})
\end{equation}
by unitarity. Therefore for present purposes it is enough to track the evolution of $\{q_{x\alpha}(t)$, $q_{z\alpha}(t)\}$ only.

Suppose one intends to entangle qubits A and D by local interactions existing only between qubits A and B, B and C, and C and D, while more distant pairs, such as A and D, are not allowed to interact directly.
In this case, $A$ and $D$ correspond to $Q_1$ and $Q_2$, while $B$ and $C$ represent the mediator $S$. The Hamiltonian of the qubits is assumed to contain nearest neighbour interactions on the chain, but not to couple qubits A and D directly. We choose a representation such that the initial conditions are expressed as $q_{z\A}(t_0)=Z\otimes id^{\otimes 3}\equiv q_{z\A}$, where $Z$ is a Pauli matrix, and so on. We choose the Heisenberg state to be $\rho_H=\ket{{\bf 0}}\bra{{\bf 0}}$, the +1 eigenstate of the operator $\frac{1}{2}(id+Z)^{\otimes 4}$.

\begin{figure}
\includegraphics{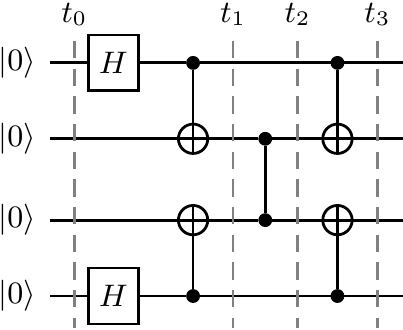}
\caption{A symmetric quantum network generating entanglement between pairs of distant qubits.}
\label{circuit1}
\end{figure}

A symmetric way of performing the maximally entangling gate between A and D is represented by the circuit in Fig.~\ref{circuit1}.
First one applies a Bell gate between A and B, and between D and C; then one performs a controlled phase on qubits B and C; and finally one applies \CNOT\ gates between A and B, and D and C.
The gates applied at their respective times are represented as follows:
\begin{equation}
\CNOT_{\alpha,\beta}(t_n)= {\textstyle \frac{1}{2}} \left (id+q_{z\alpha}(t_n)\right )+{\textstyle \frac{1}{2}} \left (id-q_{z\alpha}(t_n)\right )q_{x\beta}(t_n)
\end{equation}
\begin{equation}
\textsc{cph}_{\alpha,\beta}(t_n)= {\textstyle \frac{1}{2}} \left (id+q_{z\alpha}(t_n)\right )+{\textstyle \frac{1}{2}} \left (id-q_{z\alpha}(t_n)\right )q_{z\beta}(t_n)
\end{equation}
\begin{equation}
H_{\alpha}(t_n)= \frac{1}{\sqrt{2}} \left (q_{z\alpha}(t_n)+q_{x\alpha}(t_n)\right)
\end{equation}
The resulting evolution of the Heisenberg descriptors $\{q_{x\alpha}(t)$, $q_{z\alpha}(t)\}$ at each of the four times indicated in the figure is shown in Table 1.
\begin{table*}[htb]
\caption{Heisenberg Picture Representation---symmetric network. For each time $t$, the first slot is the $x$ component $q_{x\alpha}(t)$, the second slot is the $z$ component $q_{z\alpha}(t)$, expressed as a function of the descriptors at time $t_0$.}
\centering
\begin{tabular}{c c c c c}
\hline\hline
&Qubit A & Qubit B & Qubit C & Qubit D \\ [0.5ex] % inserts table %heading
\hline
$t_0$&$\{q_{x\A}, q_{z\A}\}$ &$\{q_{x\B}, q_{z\B}\} $&$\{q_{x\C}, q_{z\C} \}$&$\{q_{x\D}, q_{z\D}\}$\\
$t_1$&$\{q_{z\A}q_{x\B}, q_{x\A}\}$ &$\{q_{x\B}, q_{z\B}q_{x\A}\} $&$\{q_{x\C}, q_{z\C}q_{x\D} \}$&$\{q_{z\D}q_{x\C}, q_{x\D}\}$\\
$t_2$&$\{q_{z\A} q_{x\B},q_{x\A}\}$ &$\{q_{x\B}q_{z\C}q_{x\D}, q_{z\B}q_{x\A}\} $&$\{q_{x\C}q_{z\B}q_{x\A}, q_{x\D}q_{z\C} \}$&$\{q_{z\D}q_{x\C}, q_{x\D}\}$\\
$t_3$&$\{q_{z\A}q_{z\C}q_{x\D}, q_{x\A}\}$ &$\{q_{x\B}q_{z\C}q_{x\D}, q_{z\B}\} $&$\{q_{x\C}q_{z\B}q_{x\A}, q_{z\C} \}$&$\{q_{z\B}q_{z\D}q_{x\A}, q_{x\D}\}$\\
[1ex]
\hline
\end{tabular}
\label{table:HEI}
\end{table*}

From this table, one can compute the {\sl degree of entanglement} between qubits A and D after time $t_3$. The most straightforward measure to use is the sum of correlations in two complementary directions written as
\begin{equation}
E_{AD}={\rm Tr}\{\rho_H (q_{x\A}(t_3)q_{z\D}(t_3)+q_{z\A}(t_3)q_{x\D}(t_3))\}\;\label{EQ1}
\end{equation}
which for the this network has a value of $2$. Disentangled states cannot exceed the value of 1 as far as this observable is concerned (which therefore also makes it a useful entanglement witness in more general contexts, very closely related to Bell's inequalities).

By tracking the evolution of the descriptors in the table one can see the explicit role of {\sl locality of interactions}, which couple $Q_1$ and $Q_2$ separately with the mediator $S$, but not $Q_1$ and $Q_2$ directly \cite{Marletto2017,Bose2017,Bose2019}. The descriptors of system D at time $t_3$ (representing $Q_2$ in the simulation) become dependent on the descriptors of A (representing $Q_1$ in the simulation) at time $t_0$, and similarly for the descriptors of A at time $t_3$, via a sequence of nearest-neighbour interactions.

Now we can relate the final degree of entanglement between A and D to the degree of non-classicality of the mediator. A and D become entangled at time $t_3$ because at time $t_1$ and $t_2$ the non-commuting degrees of freedom $q_{z\B}$, $q_{x\B}$, and  $q_{z\C}$, $q_{x\C}$, of qubits B and C have acted as mediators. The relevant {\sl degree of non-classicality} of the mediator represented by each of the qubits B and C will be taken to be the norm of the operators $[q_{x\alpha}(t),\, q_{z\alpha}(t)]$, that is, the commutator between the two observables that are relevant for the couplings between the mediator qubits B and C  and the two qubits to be entangled, A and D. This is a dynamical quantity, which is of course invariant under unitary dynamics. We shall now simulate the transition to a classical mediator via introducing decoherence on qubits B and C, which affects that degree of non-commutativity and therefore the capacity of the mediator to create entanglement.

\subsection{Decoherence}

We will now simulate the transition between the case where the mediator $S$ consists of a fully fledged two-qubit system, and the case where it consists of a hybrid system with a lower degree of non-classicality. This will be represented in our simulation by applying some decoherence to qubits B and C.  Specifically, we apply a phase-flip channel with intensity $p$ at time $t_2$, after the phase gate and before the final \textsc{cnot} gates in the above network, to both qubits B and C separately.

We will consider the regime where the decoherence rate is faster than the timescales over which the observables of the qubits B and C can be measured. In such a situation, we can consider the mediator to be described by an effective description, where the descriptors of qubits B and C are acted upon by the noisy operation. To model this effective system (equivalent to qubit B and C each ``dressed'' by decoherence), we shall use the Heisenberg picture representation of noisy channels, where for a general observable $O_{\alpha}(t)$ of qubit $\alpha$, the phase-flip channel has the effect
\begin{equation}
E(O_{\alpha}(t))=\sum_aM_a(t)^{\dagger}O_{\alpha}(t) M_a(t)
\end{equation}
where the $M_a$ are the Kraus operators of the channel: $M_0(t)=\sqrt{p}\,id$, $M_1(t)=\sqrt{1-p}\,q_{z\alpha}(t)$.  Note that after this operation, the generators of the algebra of qubit $\alpha$ undergoing decoherence are affected as
\begin{align}
E(q_{z\alpha}(t))&= q_{z\alpha}(t),\\
E(q_{x\alpha}(t))&= (1-2p)q_{x\alpha}(t).
\end{align}

The effect of decoherence on the overall entanglement generation can be retrieved by computing the evolution of the components of the qubits, as in Table~2.
\begin{table*}[htb]
\caption{Heisenberg Picture Representation---symmetric network with decoherence.
A dephasing channel is applied to each of qubits B and C after time $t_2$.}
\centering
\begin{tabular}{c c c c c}
\hline\hline
&Qubit A & Qubit B & Qubit C & Qubit D \\ [0.5ex] % inserts table %heading
\hline
$t_0$&$\{q_{x\A}, q_{z\A}\}$ &$\{q_{x\B}, q_{z\B}\} $&$\{q_{x\C}, q_{z\C} \}$&$\{q_{x\D}, q_{z\D}\}$\\
$t_1$&$\{q_{z\A}q_{x\B}, q_{x\A}\}$ &$\{q_{x\B}, q_{z\B}q_{x\A}\} $&$\{q_{x\C}, q_{z\C}q_{x\D} \}$&$\{q_{z\D}q_{x\C}, q_{x\D}\}$\\
$t_2$&$\{q_{z\A} q_{x\B},q_{x\A}\}$ &$\{(1-2p)q_{x\B}q_{z\C}q_{x\D}, q_{z\B}q_{x\A}\} $&$\{(1-2p)q_{x\C}q_{z\B}q_{x\A}, q_{x\D}q_{z\C} \}$&$\{q_{z\D}q_{x\C}, q_{x\D}\}$\\
$t_3$&$\{(1-2p)q_{z\A}q_{z\C}q_{x\D}, q_{x\A}\}$ &$\{(1-2p)q_{x\B}q_{z\C}q_{x\D}, q_{z\B}\} $&$\{(1-2p)q_{x\C}q_{z\B}q_{x\A}, q_{z\C} \}$&$\{(1-2p)q_{z\B}q_{z\D}q_{x\A}, q_{x\D}\}$\\
[1ex]
\hline
\end{tabular}
\label{table:HEID}
\end{table*}
Computing again the degree of entanglement on qubits A and D at time $t_3$, we see that it is reduced by a factor $(1-2p)$ compared to Eqn.~\ref{EQ1}. %:

Correspondingly, the degree of commutativity of the variables of each decohered qubit involved in entanglement generation is reduced by the factor $(1-2p)$ compared to the case without decoherence:
\begin{equation}
[E(q_{z\alpha}(t)),E(q_{x\alpha}(t))]= (1-2p)[q_{z\alpha}(t), q_{x\alpha}(t)]\;.
\end{equation}

This is the dual of the channel's action on quantum states, which need not preserve the inner product between two generic quantum states. The mediator $S$ consisting of qubits B and C together with the environment that decoheres both of the qubits, can be effectively described as a physical system whose descriptors are the decohered versions of the generators of qubits B and C. This can be seen as a special case of a hybrid quantum-classical dynamics, closely related to the proposals reviewed in \cite{Sherry1979}.

In the limiting case of complete dephasing, $p=1/2$, the entanglement vanishes and so does the degree of non-commutativity between the different components of the mediator consisting of the fully decohered mediator qubits B and C, thereby simulating the transition of the mediator to a completely classical system. The physical interpretation of this is that in the limit of maximal decoherence the qubits mediating the interaction become effectively two classical bits, in that only their $z$ component can be used for information processing. Their $x$ and $y$ components become effectively suppressed, and the qubits are turned into effectively classical systems with only one classical Boolean observable, along $z$.

This is a powerful illustration of the fact that the mediator $S$ must have at least two non-commuting observables to mediate entanglement between $A$ and $D$: if the degree of commutativity (i.e.~non-classicality) of the relevant variables mediating entanglement is reduced, the final entanglement is too. Of course the mediator in this quantum simulation obeys the laws of quantum theory, but this scenario serves to illustrate the more general principle of the argument for non-classicality, which could also apply to systems that are not necessarily quantum, e.g.\ quantum gravity. Note also that having entanglement created in this way is a sufficient condition for non-classicality of the mediator. Not having entanglement, on the other hand, may or may not imply the classicality of the mediator.
One important difference between this model and the field-theory model, when the mediator is, for example, gravity, is that in the linearised Hamiltonian the interaction between the qubits is weak---it cannot be modelled as a Bell gate. But this is only a superficial difference and that interaction, despite its weakness, still leads to maximal entanglement at the end, as explained in \cite{Bose2017,Marletto2017}.

\subsection{An asymmetric equivalent formulation}
\begin{figure}
\includegraphics{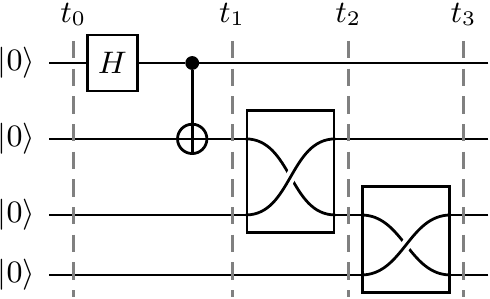}
\caption{An asymmetric quantum network transferring the entanglement between a neighbouring pair of qubits to a distant pair of qubits.}
\label{circuit2}
\end{figure}

In the next section, we will illustrate the idea of the test with an experimental simulation in an NMR spin system \cite{EBWbook,Jones2011} with four qubits.
For this simulation it is more convenient to  use an alternative asymmetric discretised network, as shown in Fig.~\ref{circuit2}.  In this alternative network the qubits A, B and D are fully quantum, while the qubit C could undergo decoherence. The logic is to prepare a maximally entangled pair on the qubits A and B and then transfer the entanglement to qubit D via two swap gates acting locally on qubits B and C and qubits C and D. See Table~\ref{table:HEIS}, where this process is described in the Heisenberg picture.
\begin{table}[b]
\caption{Heisenberg Picture Representation---asymmetric network.}
\centering
\begin{tabular}{c c c c c}
\hline\hline
&Qubit A & Qubit B & Qubit C & Qubit D \\ [0.5ex] % inserts table %heading
\hline
$t_0$&$\{q_{x\A}, q_{z\A}\}$ &$\{q_{x\B}, q_{z\B}\} $&$\{q_{x\C}, q_{z\C} \}$&$\{q_{x\D}, q_{z\D}\}$\\
$t_1$&$\{q_{z\A}q_{x\B}, q_{x\A}\}$ &$\{q_{x\B}, q_{z\B}q_{x\A}\} $&$\{q_{x\C}, q_{z\C}\}$&$\{q_{x\D}, q_{z\D}\}$\\
$t_2$&$\{q_{z\A}q_{x\B}, q_{x\A}\}$ &$\{q_{x\C}, q_{z\C} \}$&$\{q_{x\B}, q_{z\B}q_{x\A}\}$&$\{q_{x\D}, q_{z\D}\}$\\
$t_3$&$\{q_{z\A}q_{x\B}, q_{x\A}\}$ &$\{q_{x\C}, q_{z\C} \}$&$\{q_{x\D}, q_{z\D}\}$& $\{q_{x\B}, q_{z\B}q_{x\A}\}$\\
[1ex]
\hline
\end{tabular}
\label{table:HEIS}
\end{table}

At the end of the process, in the absence of decoherence, qubits A and D are maximally entangled (i.e., at time $t_3$ the witness in Eqn.~\ref{EQ1} has value 2).
In the presence of dephasing the transfer of entanglement does not happen, because the dephasing causes a progressive reduction of the non-classicality of the system C. This achieves exactly what the fully symmetric scenario we described earlier does, but is more conducive to the experiments we performed with NMR. In this network, it is enough to have one qubit ``classicalised'' in order to prevent establishment of entanglement across the chain, thus making it easier to realise via NMR simulation. More specifically, in the simulation we realise a version of the above network where the two \SWAP\ gates are achieved through a sequence of $n$ partial \SWAP\ gates, which can potentially be interrupted by decoherence on qubit C. Such decoherence is easily simulated by applying $Z$ gates probabilistically with $p=0$ for the simple quantum case and $p=1/2$ for the completely dephased classical case.

NMR systems operate at room temperature, which is high in comparison with the energy gap between spin states, and so NMR spin states are normally highly mixed, with only a small excess population in the lower energy state.  Preparation of a pure initial NMR state is only possible in special cases \cite{Anwar2004}, and instead most quantum information processing (QIP) experiments are performed using pseudo-pure states  \cite{Cory1997,Cory1998a}, or effective pure states \cite{Gershenfeld1997,Knill1998} of the form
\begin{equation}
\rho=(1-\epsilon)\textbf{1}+\epsilon|\textbf{0}\rangle\langle\textbf{0}|
\end{equation}
where $\textbf{1}$ is the maximally mixed state of the spin system and $|\textbf{0}\rangle$ is the desired initial state. As the maximally mixed state does not evolve under unitary transformations and is not detectable in NMR experiments, pseudo-pure states behave exactly like pure states except that the signal intensity is reduced to a fraction $\epsilon$.

The poor scaling of this signal intensity with the number of qubits means that conventional NMR cannot be used to perform QIP with large numbers of qubits \cite{Warren1997}. More fundamentally, the absence of entanglement in thermal pseudo-pure states has led some authors to question whether NMR devices are really quantum at all \cite{Braunstein1999}, implying that NMR experiments may be only simulations of simulations. It has, however, proved impossible to develop a fully classical model of NMR QIP \cite{Schack1999}, and NMR quantum computations run using entangled states produce identical results to those with pseudo-pure states with the exception of the increased signal size \cite{Anwar2004b}.

A second issue which arises in NMR QIP is the absence of true projective measurements, as the NMR signal detection corresponds to ensemble averaged weak measurements. This makes a convincing NMR implementation of the symmetric circuit (Fig.~\ref{circuit1}) challenging, but in the case of the asymmetric circuit (Fig.~\ref{circuit2}) a particularly simple observation scheme can be used which immediately demonstrates the transfer or otherwise of the entangled state along the chain.

\section{Experimental simulation}
\begin{figure}[tb]
\centering
\includegraphics[width=85mm]{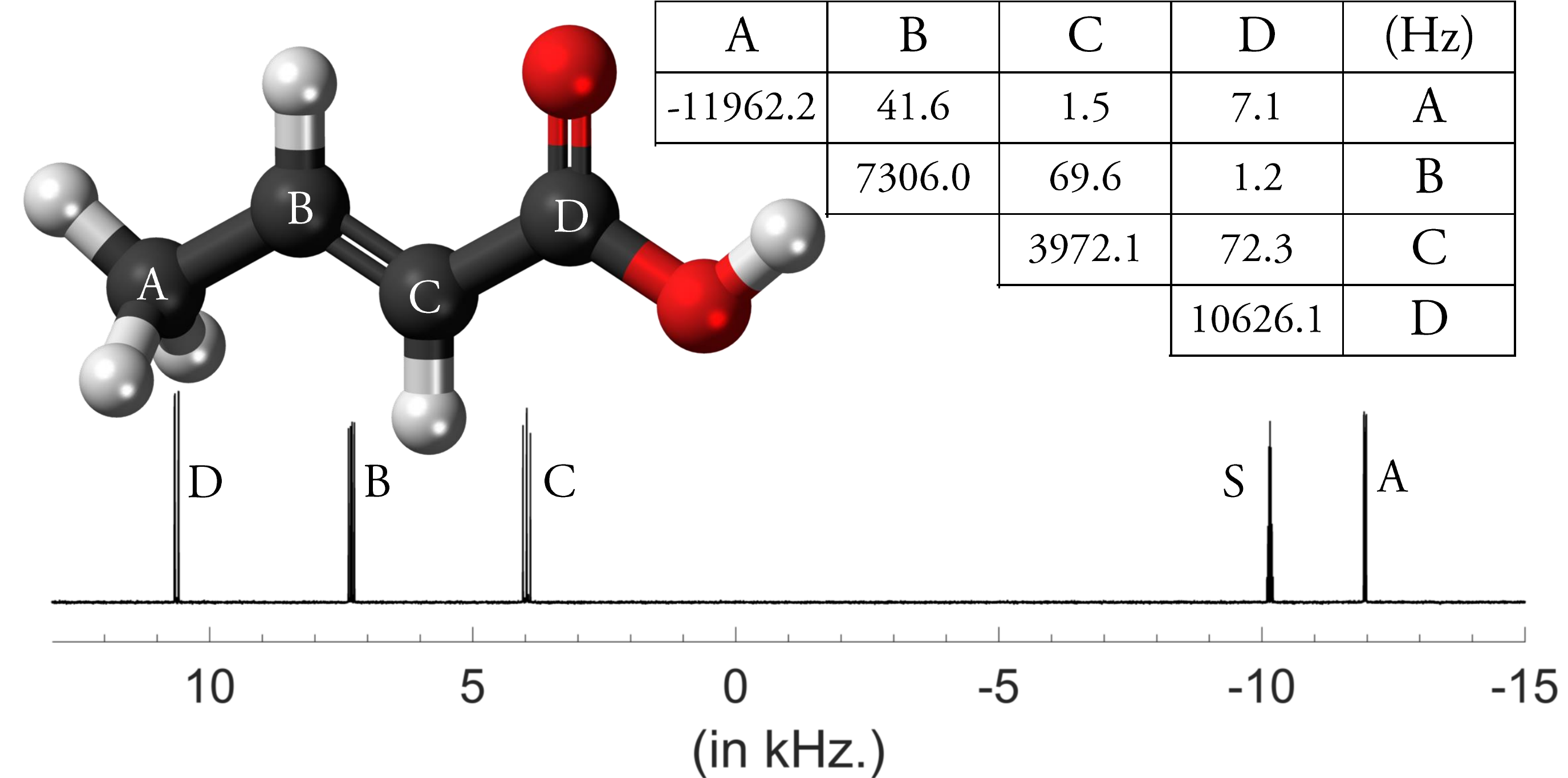}
\caption{The molecular structure and Hamiltonian parameters (offset frequencies and spin--spin couplings) for \nuc{13}{C} labelled crotonic acid. The multiplet labelled S comes from the solvent, deuterated acetone. Measured $\mathrm{T_{2}}$ relaxation times were around 1.3\,s for each spin, while $\mathrm{T_{1}}$ varied between 10\,s for spin A and 22\,s for spin D.}
\label{ham}
\end{figure}

\begin{figure}
\includegraphics{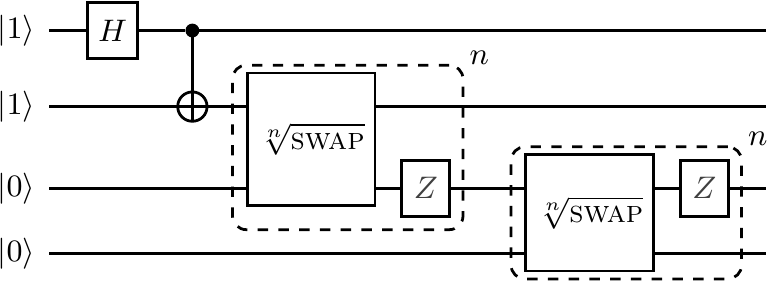}
\caption{The quantum network for the simulation. The Z gates shown in grey are applied probabilistically to simulate decoherence, as described in the main text. The gates in each dashed box are repeated $n$ times.}
\label{circuit3}
\end{figure}

Our experimental qubits are the four \nuc{13}{C} nuclei in fully labelled crotonic acid dissolved in deuterated acetone \cite{Boulant2002} at 300\,K, as shown in Fig.~\ref{ham}.
Experiments were performed on a 600\,MHz Varian Unity Inova spectrometer.
The \nuc{13}{C} NMR spectrum of the thermal equilibrium state with \nuc{1}{H} decoupling, shown at the bottom, shows that the four multiplets (one from each spin as indicated) are well separated and so can be individually addressed. The spin system can be approximated by a linear chain, with strong nearest-neighbour couplings and much weaker next-nearest-neighbour couplings. Fortuitously the long-range AD coupling is large enough to be easily resolved, rendering the detection of long-range entanglement straightforward. The circuit implemented in our experiment is shown in Fig.~\ref{circuit3}.
 NMR experiments with this spin system have recently been used \cite{Li2019} to simulate spinfoam vertex amplitudes for loop quantum gravity.

\subsection{Initial state preparation}
A pseudo-pure initial state (PPS) was prepared by spatial averaging following the methods in \cite{Sharf2000,Kong2017}. The use of robust GRAPE pulses, as described below, was found to almost double the observed signal intensity, which was then enhanced still further using the nuclear Overhauser effect from \nuc{1}{H} nuclei to generate a non-thermal initial state with enhanced polarization \cite{Jones2000a}. We chose the initial state $\ket{1100}$ as this leads to a singlet entangled state, which is the most robust of the four Bell states to naturally occurring decoherence processes.

\subsection{Robust GRAPE pulses}
All the gates used in the PPS preparation sequence as well as the quantum circuit were implemented using GRAPE \cite{Khaneja2005}.  In addition to four \nuc{13}{C} nuclei (the system qubits) crotonic acid also contains five \nuc{1}{H} nuclei (environment qubits) which have strong interactions with the system qubits. (The remaining three nuclei comprise two \nuc{16}{O} nuclei, which are spin-0 and so can be safely ignored, and a sixth \nuc{1}{H} nucleus in the hydroxyl group which undergoes rapid chemical exchange, averaging out its interactions with the main spins \cite{EBWbook}). The traditional approach is to apply \nuc{1}{H} decoupling to the environment qubits throughout the experiment, usually with a composite pulse based broadband decoupling sequence such as WALTZ-16 \cite{Shaka1983a}.  In principle such decoupling can completely trace out the environment qubits, leaving a simple four-spin system.  In practice, however, it is not possible to achieve completely effective decoupling without using high RF powers which are ruled out by hardware limits and the effects of sample heating.  Our simulations suggest that imperfect \nuc{1}{H} decoupling is the main source of errors in current NMR implementations of GRAPE pulses in crotonic acid.

Here we adopt a quite different approach: leaving the \nuc{1}{H} nuclei untouched throughout the pulse sequence which implements quantum logic gates, and applying decoupling only during the final detection stage.  In this case the \nuc{1}{H} nuclei are entirely passive, and can be thought of as providing a fixed frequency shift, which is different in each molecule, depending on the hydrogen spin states in that particular molecule. As there are five hydrogen spins there are $2^5=32$ possible spin states, although these give rise to only 16 distinct frequency shifts, as the three hydrogens attached to carbon A, forming a methyl group, are completely equivalent.  The GRAPE pulses are then designed by optimising for all 16 background Hamiltonians simultaneously, defining the overall fidelity as the average of the individual fidelities \cite{Khaneja2005}.

This approach completely avoids errors arising from imperfect decoupling, as decoupling is not applied during GRAPE pulses or free evolution periods.  As usual it is possible to design pulses which can tolerate the RF amplitude inhomogenity over the macroscopic sample, by evaluating the fidelity over a range of RF amplitudes.  Although preparing such GRAPE pulses is computationally expensive, one can greatly speed it up by employing subsystem methods \cite{Ryan2008} and using GRAWME \cite{Bhole2018} during the initial stages of the optimisation.

\subsection{SWAP gates and dephasing}
The \textsc{swap} gates were implemented in eight stages using
\begin{equation}
U_{BC}=\sqrt[8]{\textsc{swap}_{BC}}
\end{equation}
and similarly for $U_{CD}$. Poor experimental results are obtained if the same GRAPE pulse is applied repeatedly, as the small errors which inevitably occur in any experimental implementation build up linearly on repeated application \cite{Murphy2019}. Instead multiple GRAPE pulses were designed for each gate, by using different random starting points in the GRAPE search.  As each pulse has different implementation errors the total error will only grow with the square-root of the number of gates, leading to visibly better results.

Dephasing can be implemented using either spatial averaging, using magnetic field gradients \cite{Cory1997}, or by temporal averaging \cite{Knill1998}, in which spectra are recorded both with and without the application of $Z$ gates at any given point and the results combined.  Temporal averaging has the advantage that it is much simpler to implement selective dephasing on a single qubit, but if performed naively requires $2^n$ separate experiments, where dephasing is applied at $n$ separate points, rendering such experiments infeasible in all but the simplest cases \cite{Kawamura2010}.

Instead we adopted the method of randomized temporal averaging \cite{Knill1998}, in which exhaustive averaging is replaced by averaging over a sample of possible sequences. Specifically, we designed two further gates, $V_{BC}=Z_CU_{BC}$ and $V_{CD}=Z_CU_{CD}$, and used a randomly chosen sequence of $U$ and $V$ gates, containing four of each.  When the number of stages is small, as used here, the final result depends on the precise pattern of gates used, and so results were averaged over 16 different dephasing patterns.

\subsection{Detection}
Entangled states are not directly detectable in NMR: the observed signal depends on the expectation values of single-spin off-diagonal operators \cite{Goldman1988}, and these are all zero for entangled states. Here, however, we use an indirect witness \cite{Filgueiras2012}, which corresponds effectively to measuring the observable in Eqn.~\ref{EQ1}.

Consider the initial entangled state of qubits A and B, and the effect of performing a Hadamard operation on qubit A before observing the NMR spectrum. This will generate a pair of lines in the multiplet of transitions of spin A, one with positive intensity and another with negative intensity, called an antiphase doublet. However it also generates another antiphase doublet in the multiplet of spin B, even though spin B was not directly excited. There are no signals in the multiplets corresponding to spins C or D. In general, if we have an entangled singlet state between two qubits then applying a Hadamard gate to one of them generates antiphase doublets in both multiplets, so we can very easily follow the progress of the entangled state from AB to AD by applying a Hadamard to qubit A and observing the signal on spins B, C and D.

\subsection{Experimental results}
\begin{figure}[tb]
\centering
\includegraphics[width=85mm]{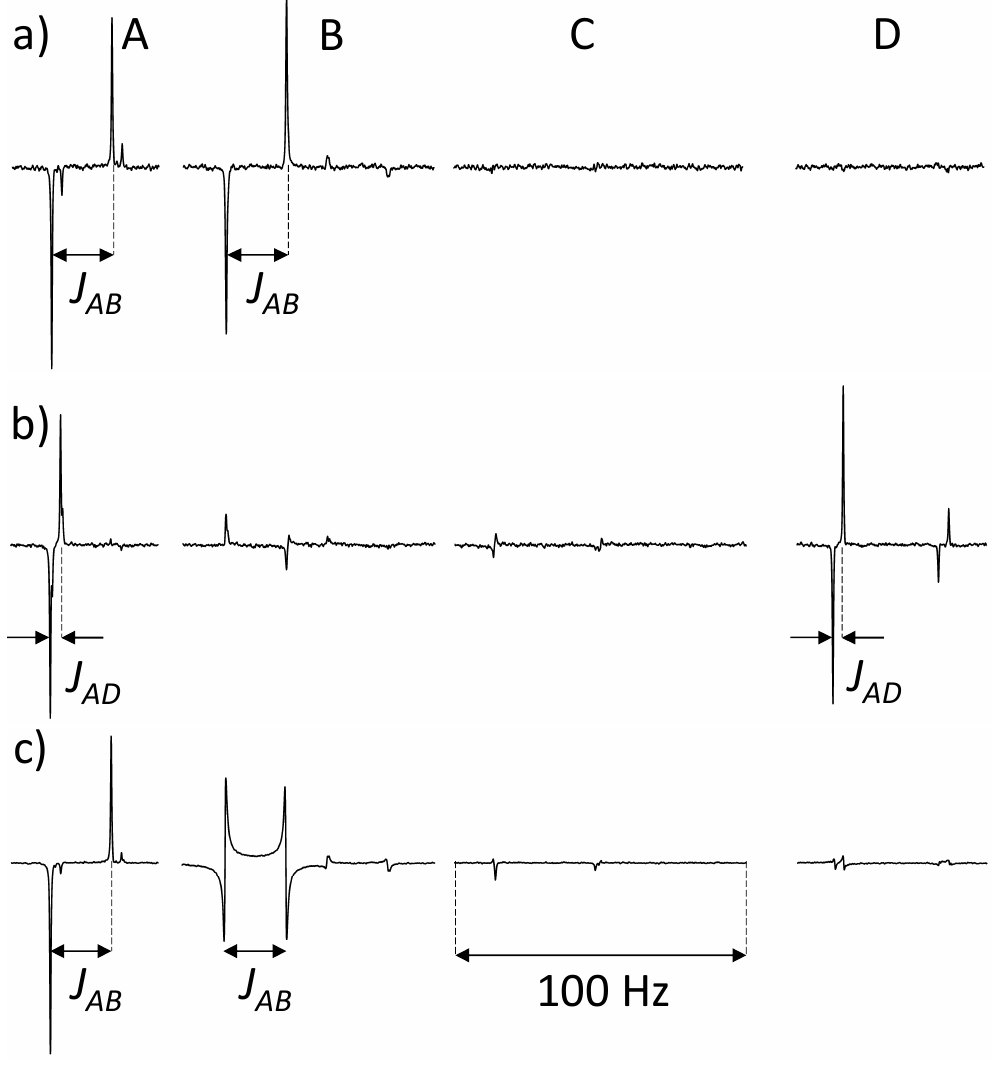}
\caption{Experimental spectra of the \nuc{13}{C} spins (A, B, C, D): a) in the initial singlet state; b) after the \textsc{swap} gates; c) after the \textsc{swap} gates interrupted by dephasing on spin C. All spectra are plotted on the same vertical scale, and the horizontal frequency scale is as indicated. The higher signal-to-noise ratio in c) reflects the averaging over 16 different dephasing patterns.}
\label{spectra}
\end{figure}

The experimental results are shown in Fig.~\ref{spectra}. Each spectrum is obtained after applying a Hadamard gate to spin A with a two step phase cycle to reduce errors \cite{Jones2011}; the \nuc{1}{H} environment nuclei were decoupled throughout signal acquisition. Individual multiplets were then cut out of spectra like that shown in Fig.~\ref{ham}, permitting an expanded horizontal scale, and then rearranged into spin order. Antiphase signal on both spins $r$ and $s$  with a coupling $J_{rs}$ indicates that spins $r$ and $s$ were entangled; smaller peaks are due to remaining experimental errors.

An AB singlet state was initially prepared, as shown by the antiphase doublets split by $J_{AB}$ on both spins A and B in spectrum (a). The transfer of entanglement from spins A and B to spins A and D is clearly seen in (b), while dephasing suppresses this transfer as shown in (c). This suppression can be seen as a generalisation of the quantum Zeno effect in NMR \cite{Xiao2006a} where rapid dephasing suppresses coherent evolution.  The phase shift on spin B in spectrum (c) arises from the $zz$ component  of the \textsc{swap} Hamiltonian which commutes with the dephasing process and so is not suppressed by the Zeno effect.

The small additional peaks are due to experimental errors, and simulations suggest that these can be largely modelled by weak uniform depolarisation during the \SWAP\ process.  This apparent depolarisation arises from the accumulation of small errors in the $U$ and $V$ gates, which can be treated as random because of the way in which particular implementations of these gates are selected from a larger set.

\section{Conclusions}
We have illustrated, with our theoretical analysis and the experimental simulation, the relevance of non-commuting degrees of freedom in the physical system mediating an entangling gate between two spatially separated qubits. The capacity to generate entanglement can thus be used as an indirect witness of non-classicality in physical systems that need not obey quantum theory, such as a macroscopic system or the gravitational field \cite{MAR}. Future applications of this general scheme will include exploring experimental schemes to witness non-classicality in bio-molecules, including living systems, which are notorious for being hard to manipulate directly, but can easily be accessed by quantum probes \cite{PAT}. Also, a worthwhile future experimental direction is to probe different regimes of decoherence and their effect on entanglement transfer.

\begin{acknowledgments}
GB is supported by a Felix Scholarship. CM thanks the Templeton World Charity Foundation and the Eutopia Foundation. VV's research is supported by the National Research Foundation, Prime Minister's Office, Singapore, under its Competitive Research Programme (CRP Award No. NRF-CRP14-2014-02) and administered by Centre for Quantum Technologies, National University of Singapore. Quantum circuits were drawn using Quantikz \cite{Kay2018}.
\end{acknowledgments}

\bibliography{ref}

\begin{thebibliography}{36}
\expandafter\ifx\csname natexlab\endcsname\relax\def\natexlab#1{#1}\fi
\expandafter\ifx\csname bibnamefont\endcsname\relax
  \def\bibnamefont#1{#1}\fi
\expandafter\ifx\csname bibfnamefont\endcsname\relax
  \def\bibfnamefont#1{#1}\fi
\expandafter\ifx\csname citenamefont\endcsname\relax
  \def\citenamefont#1{#1}\fi
\expandafter\ifx\csname url\endcsname\relax
  \def\url#1{\texttt{#1}}\fi
\expandafter\ifx\csname urlprefix\endcsname\relax\def\urlprefix{URL }\fi
\providecommand{\bibinfo}[2]{#2}
\providecommand{\eprint}[2][]{\url{#2}}

\bibitem[{\citenamefont{Marletto and
  Vedral}(2017{\natexlab{a}})}]{Marletto2017}
\bibinfo{author}{\bibfnamefont{C.}~\bibnamefont{Marletto}} \bibnamefont{and}
  \bibinfo{author}{\bibfnamefont{V.}~\bibnamefont{Vedral}},
  \bibinfo{journal}{Phys. Rev. Lett.} \textbf{\bibinfo{volume}{119}},
  \bibinfo{pages}{240402} (\bibinfo{year}{2017}{\natexlab{a}}).

\bibitem[{\citenamefont{Bose et~al.}(2017)\citenamefont{Bose, Mazumdar, Morley,
  Ulbricht, Toro\ifmmode~\check{s}\else \v{s}\fi{}, Paternostro, Geraci,
  Barker, Kim, and Milburn}}]{Bose2017}
\bibinfo{author}{\bibfnamefont{S.}~\bibnamefont{Bose}},
  \bibinfo{author}{\bibfnamefont{A.}~\bibnamefont{Mazumdar}},
  \bibinfo{author}{\bibfnamefont{G.~W.} \bibnamefont{Morley}},
  \bibinfo{author}{\bibfnamefont{H.}~\bibnamefont{Ulbricht}},
  \bibinfo{author}{\bibfnamefont{M.}~\bibnamefont{Toro\ifmmode~\check{s}\else
  \v{s}\fi{}}}, \bibinfo{author}{\bibfnamefont{M.}~\bibnamefont{Paternostro}},
  \bibinfo{author}{\bibfnamefont{A.~A.} \bibnamefont{Geraci}},
  \bibinfo{author}{\bibfnamefont{P.~F.} \bibnamefont{Barker}},
  \bibinfo{author}{\bibfnamefont{M.~S.} \bibnamefont{Kim}}, \bibnamefont{and}
  \bibinfo{author}{\bibfnamefont{G.}~\bibnamefont{Milburn}},
  \bibinfo{journal}{Phys. Rev. Lett.} \textbf{\bibinfo{volume}{119}},
  \bibinfo{pages}{240401} (\bibinfo{year}{2017}).

\bibitem[{\citenamefont{Marletto and
  Vedral}(2017{\natexlab{b}})}]{Marletto2017a}
\bibinfo{author}{\bibfnamefont{C.}~\bibnamefont{Marletto}} \bibnamefont{and}
  \bibinfo{author}{\bibfnamefont{V.}~\bibnamefont{Vedral}},
  \bibinfo{journal}{npj Quantum Information} \textbf{\bibinfo{volume}{3}},
  \bibinfo{pages}{29} (\bibinfo{year}{2017}{\natexlab{b}}).

\bibitem[{\citenamefont{Marletto and
  Vedral}(2017{\natexlab{c}})}]{Marletto2017b}
\bibinfo{author}{\bibfnamefont{C.}~\bibnamefont{Marletto}} \bibnamefont{and}
  \bibinfo{author}{\bibfnamefont{V.}~\bibnamefont{Vedral}},
  \bibinfo{journal}{npj Quantum Information} \textbf{\bibinfo{volume}{3}},
  \bibinfo{pages}{41} (\bibinfo{year}{2017}{\natexlab{c}}).

\bibitem[{\citenamefont{Deutsch and Hayden}(2000)}]{DEU}
\bibinfo{author}{\bibfnamefont{D.}~\bibnamefont{Deutsch}} \bibnamefont{and}
  \bibinfo{author}{\bibfnamefont{P.}~\bibnamefont{Hayden}},
  \bibinfo{journal}{Proc. R. Soc. Lond. A} \textbf{\bibinfo{volume}{456}},
  \bibinfo{pages}{1756} (\bibinfo{year}{2000}).

\bibitem[{\citenamefont{Gottesman}(1999)}]{Gottesman1999}
\bibinfo{author}{\bibfnamefont{D.}~\bibnamefont{Gottesman}},
  \emph{\bibinfo{title}{Proceedings of the XXII International Colloquium on
  Group Theoretical Methods in Physics}} (\bibinfo{publisher}{International
  Press}, \bibinfo{year}{1999}), chap. \bibinfo{chapter}{The Heisenberg
  Representation of Quantum Computers}, pp. \bibinfo{pages}{32--43},
  \bibinfo{note}{arXiv preprint arXiv:quant-ph/9807006}.

\bibitem[{\citenamefont{Marshman et~al.}(2019)\citenamefont{Marshman, Mazumdar,
  and Bose}}]{Bose2019}
\bibinfo{author}{\bibfnamefont{R.~J.} \bibnamefont{Marshman}},
  \bibinfo{author}{\bibfnamefont{A.}~\bibnamefont{Mazumdar}}, \bibnamefont{and}
  \bibinfo{author}{\bibfnamefont{S.}~\bibnamefont{Bose}},
  \bibinfo{journal}{arXiv preprint arXiv:arXiv:1907.01568}
  (\bibinfo{year}{2019}).

\bibitem[{\citenamefont{Sherry and Sudarshan}(1979)}]{Sherry1979}
\bibinfo{author}{\bibfnamefont{T.~N.} \bibnamefont{Sherry}} \bibnamefont{and}
  \bibinfo{author}{\bibfnamefont{E.~C.~G.} \bibnamefont{Sudarshan}},
  \bibinfo{journal}{Phys. Rev. D} \textbf{\bibinfo{volume}{20}},
  \bibinfo{pages}{857} (\bibinfo{year}{1979}).

\bibitem[{\citenamefont{Ernst et~al.}(1987)\citenamefont{Ernst, Bodenhausen,
  and Wokaun}}]{EBWbook}
\bibinfo{author}{\bibfnamefont{R.~R.} \bibnamefont{Ernst}},
  \bibinfo{author}{\bibfnamefont{G.}~\bibnamefont{Bodenhausen}},
  \bibnamefont{and} \bibinfo{author}{\bibfnamefont{A.}~\bibnamefont{Wokaun}},
  \emph{\bibinfo{title}{Principles of Nuclear Magnetic Resonance in One and Two
  Dimensions}} (\bibinfo{publisher}{Oxford University Press},
  \bibinfo{year}{1987}).

\bibitem[{\citenamefont{Jones}(2011)}]{Jones2011}
\bibinfo{author}{\bibfnamefont{J.~A.} \bibnamefont{Jones}},
  \bibinfo{journal}{Prog. NMR Spectrosc.} \textbf{\bibinfo{volume}{59}},
  \bibinfo{pages}{91} (\bibinfo{year}{2011}).

\bibitem[{\citenamefont{Anwar et~al.}(2004{\natexlab{a}})\citenamefont{Anwar,
  Blazina, Carteret, Duckett, Halstead, Jones, Kozak, and Taylor}}]{Anwar2004}
\bibinfo{author}{\bibfnamefont{M.~S.} \bibnamefont{Anwar}},
  \bibinfo{author}{\bibfnamefont{D.}~\bibnamefont{Blazina}},
  \bibinfo{author}{\bibfnamefont{H.~A.} \bibnamefont{Carteret}},
  \bibinfo{author}{\bibfnamefont{S.~B.} \bibnamefont{Duckett}},
  \bibinfo{author}{\bibfnamefont{T.~K.} \bibnamefont{Halstead}},
  \bibinfo{author}{\bibfnamefont{J.~A.} \bibnamefont{Jones}},
  \bibinfo{author}{\bibfnamefont{C.~M.} \bibnamefont{Kozak}}, \bibnamefont{and}
  \bibinfo{author}{\bibfnamefont{R.~J.~K.} \bibnamefont{Taylor}},
  \bibinfo{journal}{Phys. Rev. Lett.} \textbf{\bibinfo{volume}{93}},
  \bibinfo{pages}{040501} (\bibinfo{year}{2004}{\natexlab{a}}).

\bibitem[{\citenamefont{Cory et~al.}(1997)\citenamefont{Cory, Fahmy, and
  Havel}}]{Cory1997}
\bibinfo{author}{\bibfnamefont{D.~G.} \bibnamefont{Cory}},
  \bibinfo{author}{\bibfnamefont{A.~F.} \bibnamefont{Fahmy}}, \bibnamefont{and}
  \bibinfo{author}{\bibfnamefont{T.~F.} \bibnamefont{Havel}},
  \bibinfo{journal}{Proc. Natl. Acad. Sci. USA} \textbf{\bibinfo{volume}{94}},
  \bibinfo{pages}{1634} (\bibinfo{year}{1997}).

\bibitem[{\citenamefont{Cory et~al.}(1998)\citenamefont{Cory, Price, and
  Havel}}]{Cory1998a}
\bibinfo{author}{\bibfnamefont{D.~G.} \bibnamefont{Cory}},
  \bibinfo{author}{\bibfnamefont{M.~D.} \bibnamefont{Price}}, \bibnamefont{and}
  \bibinfo{author}{\bibfnamefont{T.~F.} \bibnamefont{Havel}},
  \bibinfo{journal}{Physica D} \textbf{\bibinfo{volume}{120}},
  \bibinfo{pages}{82} (\bibinfo{year}{1998}).

\bibitem[{\citenamefont{Gershenfeld and Chuang}(1997)}]{Gershenfeld1997}
\bibinfo{author}{\bibfnamefont{N.~A.} \bibnamefont{Gershenfeld}}
  \bibnamefont{and} \bibinfo{author}{\bibfnamefont{I.~L.}
  \bibnamefont{Chuang}}, \bibinfo{journal}{Science}
  \textbf{\bibinfo{volume}{275}}, \bibinfo{pages}{350} (\bibinfo{year}{1997}).

\bibitem[{\citenamefont{Knill et~al.}(1998)\citenamefont{Knill, Chuang, and
  Laflamme}}]{Knill1998}
\bibinfo{author}{\bibfnamefont{E.}~\bibnamefont{Knill}},
  \bibinfo{author}{\bibfnamefont{I.}~\bibnamefont{Chuang}}, \bibnamefont{and}
  \bibinfo{author}{\bibfnamefont{R.}~\bibnamefont{Laflamme}},
  \bibinfo{journal}{Phys. Rev. A} \textbf{\bibinfo{volume}{57}},
  \bibinfo{pages}{3348} (\bibinfo{year}{1998}).

\bibitem[{\citenamefont{Warren}(1997)}]{Warren1997}
\bibinfo{author}{\bibfnamefont{W.~S.} \bibnamefont{Warren}},
  \bibinfo{journal}{Science} \textbf{\bibinfo{volume}{277}},
  \bibinfo{pages}{1688} (\bibinfo{year}{1997}).

\bibitem[{\citenamefont{Braunstein et~al.}(1999)\citenamefont{Braunstein,
  Caves, Jozsa, Linden, Popescu, and Schack}}]{Braunstein1999}
\bibinfo{author}{\bibfnamefont{S.~L.} \bibnamefont{Braunstein}},
  \bibinfo{author}{\bibfnamefont{C.~M.} \bibnamefont{Caves}},
  \bibinfo{author}{\bibfnamefont{R.}~\bibnamefont{Jozsa}},
  \bibinfo{author}{\bibfnamefont{N.}~\bibnamefont{Linden}},
  \bibinfo{author}{\bibfnamefont{S.}~\bibnamefont{Popescu}}, \bibnamefont{and}
  \bibinfo{author}{\bibfnamefont{R.}~\bibnamefont{Schack}},
  \bibinfo{journal}{Phys. Rev. Lett.} \textbf{\bibinfo{volume}{83}},
  \bibinfo{pages}{1054} (\bibinfo{year}{1999}).

\bibitem[{\citenamefont{Schack and Caves}(1999)}]{Schack1999}
\bibinfo{author}{\bibfnamefont{R.}~\bibnamefont{Schack}} \bibnamefont{and}
  \bibinfo{author}{\bibfnamefont{C.~M.} \bibnamefont{Caves}},
  \bibinfo{journal}{Phys. Rev. A} \textbf{\bibinfo{volume}{60}},
  \bibinfo{pages}{4354} (\bibinfo{year}{1999}).

\bibitem[{\citenamefont{Anwar et~al.}(2004{\natexlab{b}})\citenamefont{Anwar,
  Jones, Blazina, Duckett, and Carteret}}]{Anwar2004b}
\bibinfo{author}{\bibfnamefont{M.~S.} \bibnamefont{Anwar}},
  \bibinfo{author}{\bibfnamefont{J.~A.} \bibnamefont{Jones}},
  \bibinfo{author}{\bibfnamefont{D.}~\bibnamefont{Blazina}},
  \bibinfo{author}{\bibfnamefont{S.~B.} \bibnamefont{Duckett}},
  \bibnamefont{and} \bibinfo{author}{\bibfnamefont{H.~A.}
  \bibnamefont{Carteret}}, \bibinfo{journal}{Phys. Rev. A}
  \textbf{\bibinfo{volume}{70}}, \bibinfo{pages}{032324}
  (\bibinfo{year}{2004}{\natexlab{b}}).

\bibitem[{\citenamefont{Boulant et~al.}(2002)\citenamefont{Boulant, Fortunato,
  Pravia, Teklemariam, Cory, and Havel}}]{Boulant2002}
\bibinfo{author}{\bibfnamefont{N.}~\bibnamefont{Boulant}},
  \bibinfo{author}{\bibfnamefont{E.~M.} \bibnamefont{Fortunato}},
  \bibinfo{author}{\bibfnamefont{M.~A.} \bibnamefont{Pravia}},
  \bibinfo{author}{\bibfnamefont{G.}~\bibnamefont{Teklemariam}},
  \bibinfo{author}{\bibfnamefont{D.~G.} \bibnamefont{Cory}}, \bibnamefont{and}
  \bibinfo{author}{\bibfnamefont{T.~F.} \bibnamefont{Havel}},
  \bibinfo{journal}{Phys. Rev. A} \textbf{\bibinfo{volume}{65}},
  \bibinfo{pages}{024302} (\bibinfo{year}{2002}).

\bibitem[{\citenamefont{Li et~al.}(2019)\citenamefont{Li, Li, Han, Lu, Zhou,
  Ruan, Long, Wan, Lu, Zeng et~al.}}]{Li2019}
\bibinfo{author}{\bibfnamefont{K.}~\bibnamefont{Li}},
  \bibinfo{author}{\bibfnamefont{Y.}~\bibnamefont{Li}},
  \bibinfo{author}{\bibfnamefont{M.}~\bibnamefont{Han}},
  \bibinfo{author}{\bibfnamefont{S.}~\bibnamefont{Lu}},
  \bibinfo{author}{\bibfnamefont{J.}~\bibnamefont{Zhou}},
  \bibinfo{author}{\bibfnamefont{D.}~\bibnamefont{Ruan}},
  \bibinfo{author}{\bibfnamefont{G.}~\bibnamefont{Long}},
  \bibinfo{author}{\bibfnamefont{Y.}~\bibnamefont{Wan}},
  \bibinfo{author}{\bibfnamefont{D.}~\bibnamefont{Lu}},
  \bibinfo{author}{\bibfnamefont{B.}~\bibnamefont{Zeng}}, \bibnamefont{et~al.},
  \bibinfo{journal}{Communications Physics} \textbf{\bibinfo{volume}{2}},
  \bibinfo{pages}{122} (\bibinfo{year}{2019}).

\bibitem[{\citenamefont{Sharf et~al.}(2000)\citenamefont{Sharf, Havel, and
  Cory}}]{Sharf2000}
\bibinfo{author}{\bibfnamefont{Y.}~\bibnamefont{Sharf}},
  \bibinfo{author}{\bibfnamefont{T.~F.} \bibnamefont{Havel}}, \bibnamefont{and}
  \bibinfo{author}{\bibfnamefont{D.~G.} \bibnamefont{Cory}},
  \bibinfo{journal}{Phys. Rev. A} \textbf{\bibinfo{volume}{62}},
  \bibinfo{pages}{052314} (\bibinfo{year}{2000}).

\bibitem[{\citenamefont{Kong et~al.}(2017)\citenamefont{Kong, Xin, Wei, Wang,
  Wang, Li, and Long}}]{Kong2017}
\bibinfo{author}{\bibfnamefont{X.}~\bibnamefont{Kong}},
  \bibinfo{author}{\bibfnamefont{T.}~\bibnamefont{Xin}},
  \bibinfo{author}{\bibfnamefont{S.}~\bibnamefont{Wei}},
  \bibinfo{author}{\bibfnamefont{B.}~\bibnamefont{Wang}},
  \bibinfo{author}{\bibfnamefont{Y.}~\bibnamefont{Wang}},
  \bibinfo{author}{\bibfnamefont{K.}~\bibnamefont{Li}}, \bibnamefont{and}
  \bibinfo{author}{\bibfnamefont{G.}~\bibnamefont{Long}},
  \bibinfo{journal}{arXiv preprint arXiv:1708.06050}  (\bibinfo{year}{2017}).

\bibitem[{\citenamefont{Jones}(2000)}]{Jones2000a}
\bibinfo{author}{\bibfnamefont{J.~A.} \bibnamefont{Jones}},
  \bibinfo{journal}{Fort. der Physik} \textbf{\bibinfo{volume}{48}},
  \bibinfo{pages}{909} (\bibinfo{year}{2000}).

\bibitem[{\citenamefont{Khaneja et~al.}(2005)\citenamefont{Khaneja, Reiss,
  Kehlet, Schulte-Herbr{\"u}ggen, and Glaser}}]{Khaneja2005}
\bibinfo{author}{\bibfnamefont{N.}~\bibnamefont{Khaneja}},
  \bibinfo{author}{\bibfnamefont{T.}~\bibnamefont{Reiss}},
  \bibinfo{author}{\bibfnamefont{C.}~\bibnamefont{Kehlet}},
  \bibinfo{author}{\bibfnamefont{T.}~\bibnamefont{Schulte-Herbr{\"u}ggen}},
  \bibnamefont{and} \bibinfo{author}{\bibfnamefont{S.~J.}
  \bibnamefont{Glaser}}, \bibinfo{journal}{J. Magn. Reson.}
  \textbf{\bibinfo{volume}{172}}, \bibinfo{pages}{296} (\bibinfo{year}{2005}).

\bibitem[{\citenamefont{Shaka et~al.}(1983)\citenamefont{Shaka, Keeler,
  Frenkiel, and Freeman}}]{Shaka1983a}
\bibinfo{author}{\bibfnamefont{A.~J.} \bibnamefont{Shaka}},
  \bibinfo{author}{\bibfnamefont{J.}~\bibnamefont{Keeler}},
  \bibinfo{author}{\bibfnamefont{T.}~\bibnamefont{Frenkiel}}, \bibnamefont{and}
  \bibinfo{author}{\bibfnamefont{R.}~\bibnamefont{Freeman}},
  \bibinfo{journal}{J. Magn. Reson.} \textbf{\bibinfo{volume}{52}},
  \bibinfo{pages}{335 } (\bibinfo{year}{1983}).

\bibitem[{\citenamefont{Ryan et~al.}(2008)\citenamefont{Ryan, Negrevergne,
  Laforest, Knill, and Laflamme}}]{Ryan2008}
\bibinfo{author}{\bibfnamefont{C.~A.} \bibnamefont{Ryan}},
  \bibinfo{author}{\bibfnamefont{C.}~\bibnamefont{Negrevergne}},
  \bibinfo{author}{\bibfnamefont{M.}~\bibnamefont{Laforest}},
  \bibinfo{author}{\bibfnamefont{E.}~\bibnamefont{Knill}}, \bibnamefont{and}
  \bibinfo{author}{\bibfnamefont{R.}~\bibnamefont{Laflamme}},
  \bibinfo{journal}{Phys. Rev. A} \textbf{\bibinfo{volume}{78}},
  \bibinfo{pages}{012328} (\bibinfo{year}{2008}).

\bibitem[{\citenamefont{Bhole and Jones}(2018)}]{Bhole2018}
\bibinfo{author}{\bibfnamefont{G.}~\bibnamefont{Bhole}} \bibnamefont{and}
  \bibinfo{author}{\bibfnamefont{J.~A.} \bibnamefont{Jones}},
  \bibinfo{journal}{Frontiers of Physics} \textbf{\bibinfo{volume}{13}},
  \bibinfo{pages}{130312} (\bibinfo{year}{2018}).

\bibitem[{\citenamefont{Murphy and Brown}(2019)}]{Murphy2019}
\bibinfo{author}{\bibfnamefont{D.~C.} \bibnamefont{Murphy}} \bibnamefont{and}
  \bibinfo{author}{\bibfnamefont{K.~R.} \bibnamefont{Brown}},
  \bibinfo{journal}{Phys. Rev. A} \textbf{\bibinfo{volume}{99}},
  \bibinfo{pages}{032318} (\bibinfo{year}{2019}).

\bibitem[{\citenamefont{Kawamura et~al.}(2010)\citenamefont{Kawamura, Rowland,
  and Jones}}]{Kawamura2010}
\bibinfo{author}{\bibfnamefont{M.}~\bibnamefont{Kawamura}},
  \bibinfo{author}{\bibfnamefont{B.}~\bibnamefont{Rowland}}, \bibnamefont{and}
  \bibinfo{author}{\bibfnamefont{J.~A.} \bibnamefont{Jones}},
  \bibinfo{journal}{Phys. Rev. A} \textbf{\bibinfo{volume}{82}},
  \bibinfo{pages}{032315} (\bibinfo{year}{2010}).

\bibitem[{\citenamefont{Goldman}(1988)}]{Goldman1988}
\bibinfo{author}{\bibfnamefont{M.}~\bibnamefont{Goldman}},
  \emph{\bibinfo{title}{Quantum Description of High-Resolution {NMR} in
  Liquids}} (\bibinfo{publisher}{Clarendon Press}, \bibinfo{year}{1988}).

\bibitem[{\citenamefont{Filgueiras et~al.}(2012)\citenamefont{Filgueiras,
  Maciel, Auccaise, Vianna, Sarthour, and Oliveira}}]{Filgueiras2012}
\bibinfo{author}{\bibfnamefont{J.~G.} \bibnamefont{Filgueiras}},
  \bibinfo{author}{\bibfnamefont{T.~O.} \bibnamefont{Maciel}},
  \bibinfo{author}{\bibfnamefont{R.~E.} \bibnamefont{Auccaise}},
  \bibinfo{author}{\bibfnamefont{R.~O.} \bibnamefont{Vianna}},
  \bibinfo{author}{\bibfnamefont{R.~S.} \bibnamefont{Sarthour}},
  \bibnamefont{and} \bibinfo{author}{\bibfnamefont{I.~S.}
  \bibnamefont{Oliveira}}, \bibinfo{journal}{Quant. Inf. Proc.}
  \textbf{\bibinfo{volume}{11}}, \bibinfo{pages}{1883} (\bibinfo{year}{2012}).

\bibitem[{\citenamefont{Xiao and Jones}(2006)}]{Xiao2006a}
\bibinfo{author}{\bibfnamefont{L.}~\bibnamefont{Xiao}} \bibnamefont{and}
  \bibinfo{author}{\bibfnamefont{J.~A.} \bibnamefont{Jones}},
  \bibinfo{journal}{Phys. Lett. A} \textbf{\bibinfo{volume}{359}},
  \bibinfo{pages}{424} (\bibinfo{year}{2006}).

\bibitem[{\citenamefont{Marletto and Vedral}(2018)}]{MAR}
\bibinfo{author}{\bibfnamefont{C.}~\bibnamefont{Marletto}} \bibnamefont{and}
  \bibinfo{author}{\bibfnamefont{V.}~\bibnamefont{Vedral}},
  \bibinfo{journal}{arXiv:1812.06750}  (\bibinfo{year}{2018}).

\bibitem[{\citenamefont{Krisnanda et~al.}(2018)\citenamefont{Krisnanda,
  Marletto, Vedral, Paternostro, and Paterek}}]{PAT}
\bibinfo{author}{\bibfnamefont{T.}~\bibnamefont{Krisnanda}},
  \bibinfo{author}{\bibfnamefont{C.}~\bibnamefont{Marletto}},
  \bibinfo{author}{\bibfnamefont{V.}~\bibnamefont{Vedral}},
  \bibinfo{author}{\bibfnamefont{M.}~\bibnamefont{Paternostro}},
  \bibnamefont{and} \bibinfo{author}{\bibfnamefont{T.}~\bibnamefont{Paterek}},
  \bibinfo{journal}{npj Quantum Information} \textbf{\bibinfo{volume}{4}},
  \bibinfo{pages}{60} (\bibinfo{year}{2018}).

\bibitem[{\citenamefont{Kay}(2018)}]{Kay2018}
\bibinfo{author}{\bibfnamefont{A.}~\bibnamefont{Kay}}, \bibinfo{journal}{arXiv
  preprint arXiv:1809.03842}  (\bibinfo{year}{2018}).

\end{thebibliography}

\end{document}